\documentstyle[prb,preprint,tighten,aps,graphicx]{revtex}
\begin{document}
\draft

\title{Electron Standing Wave Formation in Atomic Wires}

\author{Eldon G. Emberly\footnote{e-mail: eemberly@sfu.ca  Copyright 
Physical Review 1999} and George Kirczenow}

\address{Department of Physics, Simon Fraser University,
Burnaby, B.C., Canada V5A 1S6}

\date{\today}

\maketitle
\begin{abstract}
Using the Landauer
formulation of transport theory and tight binding models of the electronic
structure, we study electron transport through atomic wires that form
1D constrictions between pairs of metallic nano-contacts. Our results are
interpreted in terms of electron standing waves formed in the atomic wires
due to  interference of electron waves reflected at the ends of the atomic
constrictions. We explore the influence of the chemistry of the atomic
wire-metal contact interfaces on these standing waves and the associated
transport resonances by considering two types of atomic wires: gold wires
attached to gold contacts and carbon wires attached to gold contacts.  We
find that the conductance of the gold wires is roughly $1 G_0 = 2 e^2/h$
for the wire lengths studied, in agreement with experiments.  By contrast,
for the carbon wires the conductance is found to oscillate strongly as the
number of atoms in the wire varies, the odd numbered chains being
more conductive than the even numbered ones, in agreement with previous
theoretical work that was based on a different model of the carbon wire
and metal contacts.
\end{abstract}
%
\section{Introduction}
The ability to manipulate matter on the atomic scale has
improved greatly over the past decade. The STM has
facilitated much of this progress with its ability to
positionally control atoms on
surfaces.\cite{Stipe98,Weiss98,Gimz98,Andres96,Bumm96}
Experiments are now able now to move beyond the manipulation
stage and measure the electronic properties of these atomic
scale systems. One system of particular interest is an atomic
chain.  When an atomic chain is used to bridge two contacts an
atomic wire is formed.  Recent STM experiments have managed to
measure the tunneling current through atomic wires, which have
been made by positioning several atoms to form a chain between
the STM tip and the substrate.\cite{Kawa98,Yaz96} More
recently, mechanically controlled break junctions have
been used to form a gold chain between two gold contacts. The
conductance of these gold atomic wires was measured for
different chain lengths.\cite{Yan98} It may also be possible
to form carbon atomic wires by stretching a single walled
nanotube.\cite{Pier98} Current work is focusing on connecting
carbon wires between two metalized AFM tips.\cite{Yu98} Early
theoretical work on atomic wires studied the electron
transmission through molecular chains with a single orbital
per site with nearest neighbour hopping.\cite{Mujic94} Recent
theoretical studies of these atomic wire systems based on a
density functional approach have also revealed interesting
phenomena, one of them being the oscillation of conductance
versus the length of a carbon wire.\cite{Lang98}

The atomic wire is interesting from a transport point of view
because it confines electrons to propagate only in one
dimension.  It is also on a length scale that is of the order
of the wavelength of the electrons which are incident from the
contacts. This means that the quantum mechanical wave nature of
the electrons will be important. When waves are confined to
propagate in a 1D constriction standing waves are formed.
These arise because of back reflections at the boundaries
between the
contacts and the constriction. Electrons propagating in
semiconductor waveguides have been predicted to form standing
waves along the conduction channel.\cite{semiref,Ulloa92}
Experimentally observing standing wave formation in these
systems has been difficult because these structures are formed
electro-statically, which results in 1D channels that are
smoothly tapered at their ends minimizing reflections
there.\cite{adiabatic,Ulloa92} We
show that electron standing waves should form in atomic
wires and investigate the role of the chemistry of the
wire-contact interface (which has no analog in the semiconductor
devices) on their behavior.
Although these standing waves have not been studied
experimentally at the present time, atomic wire systems may
prove to be better candidates for their observation since the
interfaces between a 1D wire and the contacts cannot be smooth
on the atomic scale so that reflections at these interfaces
should be significant.

We study electron transport through an atomic wire by
calculating the wavefunction for an electron propagating along
the wire.  This is done using a tight binding formalism that
treats both the chain and metallic leads
atomistically.\cite{Ember98a} The present work is complementary
to other theoretical treatments that have been based on tight
binding models of the atomic
wire\cite{Mujic94,Samant96,Joach96} or have considered
jellium contacts with the wire modelled using
pseudo-potentials.\cite{Lang98} We apply our technique to two
different atomic wire systems. The first system is a
homogeneous atomic wire which is made using a gold chain that
bridges two gold contacts. We find that our results are
consistent with the interpretation that standing waves are
formed along the gold wire.  The standing wave patterns
that are formed are confined within the 1D channel that has the
length of the gold chain.  The number of standing wave
resonances that occurs increases with the length of the
chain. However, the conductance of gold chains consisting of
different numbers of atoms is found to vary only slightly as
the number of atoms changes and remains near unity as has been
reported experimentally.\cite{Yan98}

The second system we consider is a heterogeneous atomic wire.
It is a carbon atom chain attached to two gold contacts.  Again
our results are consistent with a description involving the
formation of standing waves along the carbon chain.  However,
in this case the standing wave patterns
formed extend significantly
beyond the ends of the chain into the gold contacts. Thus in
this heterogeneous system the chain is effectively lengthened
due to the chemistry between the chain and the contacts. This
gives rise to a greater number of standing wave resonances.
The conductance is also found to oscillate with the length of
the chain which is in agreement with what was found in other
work.\cite{Lang98} The standing wave model that we describe
provides an
intuitive physical explanation of this predicted oscillation.

In Sec. I we present the theoretical approach used to calculate the
conductance together with a description of standing
wave formation in an atomic wire.  We then present our
calculations for gold atomic wires, a homogeneous
system, in Sec. II.  In Sec. III we present calculations
for carbon atomic wires.

\section{Electron Transport in Atomic Wires: Normal Modes}

An atomic wire is composed of a chain of atoms that forms a 1D
constriction between two metallic contacts which act as the
source and the drain.  Electrons incident as Bloch waves from
the source contact scatter through the 1D constriction and
enter the drain. Landauer theory relates the conductance $G$ of
the atomic wire to the multi-channel
transmission probability $T$ for an
electron at the Fermi level of the source to scatter from
source to drain via $G = \frac{2 e^2}{h} T$.\cite{Lan57}

\begin{figure}[ht]
\includegraphics[bb = 0 0 640 800, width = 
0.85\textwidth,clip]{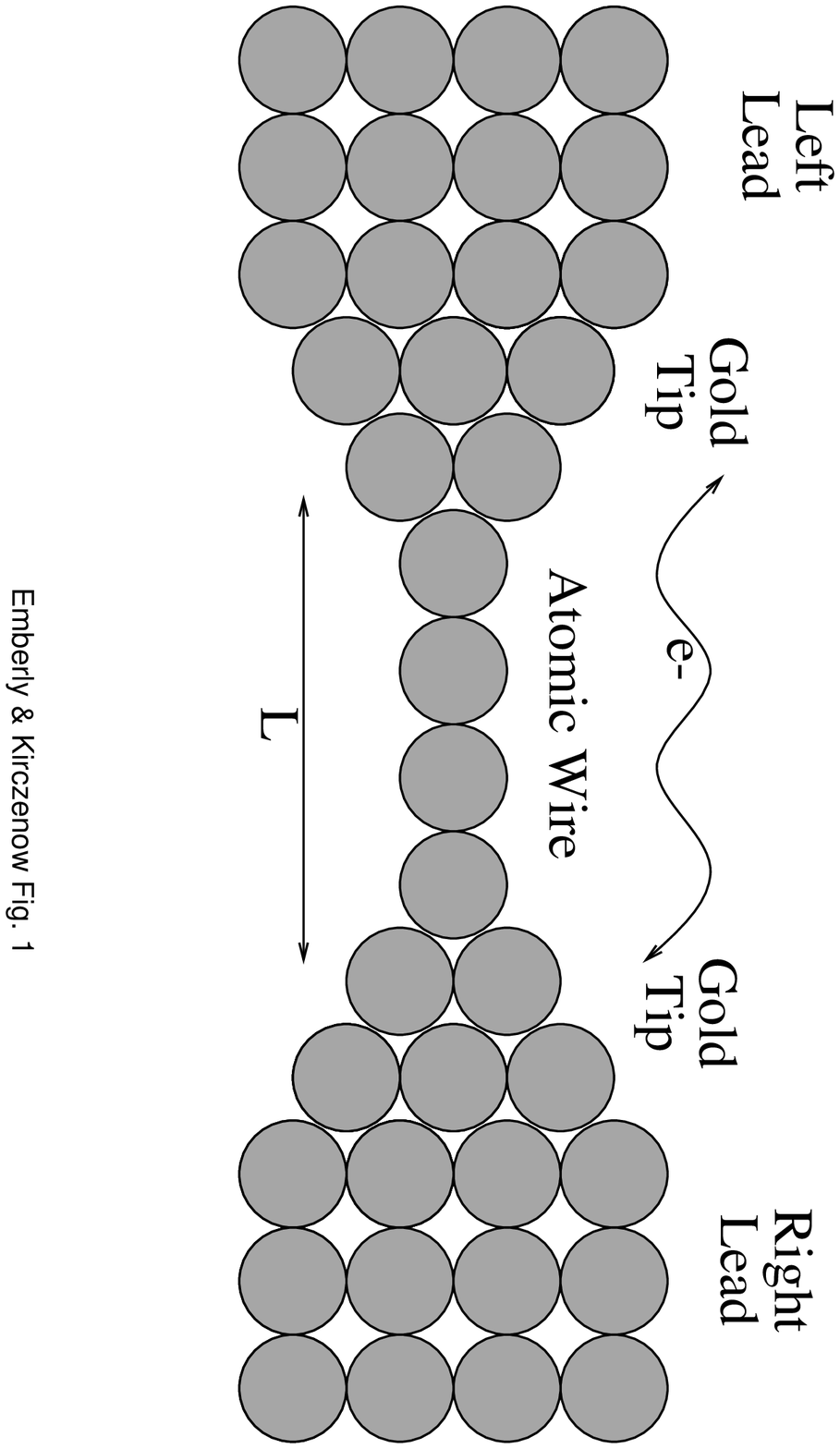} \begin{center}
\caption{Schematic diagram of 4 atom gold chain and (100) gold
tips and contacts.}
\end{center}
\label{fig1}
\end{figure}

We calculate the multi-channel
transmission probability $T$ by solving for
the ${\bf t}$ matrix of the atomic wire system.  For a given
energy $E$, an electron can propagate in one of many electronic
modes in both the source and the drain.  The ${\bf t}$ matrix
connects the modes (or channels) in the source to the channels
in the drain for a given energy. We use a tight-binding
treatment of Schr\"{o}dinger's equation to set up a system of
linear equations to solve for ${\bf t}$. The source and drain
are treated as 1D leads with multi-atom unit cells.  The wire
is attached to atomic tips which are then attached to the
leads.  An schematic diagram of an atomic wire system is shown in
Fig. \ref{fig1}.

The first step in the transport calculation is to impose
boundary conditions on the wavefunction $|\Psi^{\alpha}\rangle$
which describes an electron incident in the $\alpha^{th}$
channel with energy $E$.  In the source (or left lead (L))
$|\Psi^{\alpha}\rangle$
consists of the incident rightward propagating (+)
$\alpha^{th}$ Bloch mode $|\Phi^{\alpha}_{+}\rangle_L$ along
with a sum over reflected leftward propagating (-) and evanescent
modes $|\Phi^{\alpha'}_{-}\rangle_L$. On the atomic wire (A),
the wavefunction is a sum over the atomic orbitals
$|j\rangle$ that exist on the wire. In the drain (or right lead
(R)) the wavefunction is a sum over the transmitted
rightward propagating and evanescent modes
$|\Phi^{\alpha'}_{+}\rangle_R$.  Thus the wavefunction is given
by,
\begin{equation}
|\Psi^{\alpha}\rangle = |\Psi^{\alpha}\rangle_L
+ |\Psi^{\alpha}\rangle_A + |\Psi^{\alpha}\rangle_R
\end{equation}
where
\begin{eqnarray}
|\Psi^{\alpha}\rangle_L &=& |\Phi^{\alpha}_{+}\rangle_L +
 \sum_{\alpha'} r_{\alpha',\alpha}
|\Phi^{\alpha'}_{-}\rangle_L \\
|\Psi^{\alpha}\rangle_A &=& \sum_j d^{\alpha}_j |j\rangle \\
|\Psi^{\alpha}\rangle_R &=& \sum_{\alpha'} t_{\alpha',\alpha}
|\Phi^{\alpha'}_{+}\rangle_R.
\end{eqnarray}
Using Schr\"{o}dinger's equation $H|\Psi^{\alpha}\rangle =
E|\Psi^{\alpha}\rangle$, where $H$ is the extended H\"{u}ckel
tight
binding Hamiltonian\cite{Hoff63} of the system,
yields a system of equations which we
solve to find the $r$'s, $d$'s and $t$'s.  With the ${\bf
t}$ matrix thus determined,
the transmission probability $T$ is then
found using
\begin{equation}
T(E) = \sum_{\beta} \sum_{\alpha} \left |
\frac{v_{\beta}}{v_{\alpha}} \right | |t_{\beta,\alpha}|^2
\label{eq:multiT}
\end{equation}
where the sum over $\alpha$ is over the rightward propagating
modes in the left lead and the sum over $\beta$ is over
rightward propagating modes in the right lead.  The velocity
ratio appears since the velocities of modes in the left and
right leads may be different. The conductance is then found by
using the Landauer formula given above for the transmission
probability evaluated at the Fermi energy $E_f$ of the source.
This method and the procedure used for handling the
non-orthogonality between the different tight-binding orbitals
is described in more detail elsewhere.\cite{Ember98a,Ember98b}

The above tight-binding model is completely general and is
applicable to many types of systems, not just atomic wires.  We
now proceed to examine the physical behaviour
the above model should exhibit when applied to a 1D constriction
such as an atomic wire.  The physical interpretation that we
will use to describe the results of our transport calculations
that follow is in terms of electronic
standing waves formed within the constriction.

When waves (be they light, sound or electron) are confined to
propagate within a finite one dimensional channel, standing
waves are formed. They are formed due to the interference of
waves back reflected from the ends of the pipe where there is
a large impedance mismatch. These standing waves (also known
as normal modes) have an associated wave vector given by $k =
n \pi / L$ where $n=1,2,3,\ldots$ labels the normal mode and
$L$ is the effective length of the channel. Within the tight
binding framework, for a 1D channel
composed of a finite chain of discrete atoms equally spaced by
a distance $a$, the number of normal modes becomes finite.
The reduced wavevector of the wave is $y = k a = n \pi a / L$
and has the range $ 0 < y < \pi$. Thus $n a$ must be less than
or equal to $L$ which restricts the values that $n$ can take.
Thus a finite chain supports only a finite number of standing
waves.

In regards to electron transport, when an electron is incident
from the source with an energy that corresponds to the energy
of one of the normal modes, it will form a standing wave
along the chain.  This will correspond to maximal transmission
through the atomic wire. Because the chain supports a finite
number of normal modes the transmission probability will
have a discrete number of resonances, each one corresponding to
one of the normal modes.

The effective
length $L$ of the atomic wire determines the number of
resonances in the transmission probability.  Intuitively this
length should approximate the contact-to-contact distance.
However, an effective length which may be shorter or longer
than the contact-to-contact spacing will occur in
practice. This depends on the chemical and geometrical nature
of the interface between the atomic chain and the metallic
contacts. The chemistry may serve to effectively sharpen or
broaden the constriction which will affect the effective
length. For a homogeneous atomic wire system where the
contacts and the chain are made of the same atoms, the length
$L$ of the wire should be approximately equal to the
contact-to-contact distance. In this situation the confining
potential is that created by the geometry of the constriction
and hence the backscattering should occur at the interface
between the contacts and the chain where the impedance
mismatch is high.  For heterogeneous systems, the length $L$
may be different from the contact-to-contact length.  This is
because the chemistry between the chain and the contact is
significantly different than that within the bulk of the
contact.  Thus the confining potential due to the geometry is
now altered. An interface region is created which modifies the
confining potential and this changes the effective length $L$
of the chain.  So in heterogeneous wires one may expect there
to be more or fewer resonances than in the homogeneous case.

\section{Homogeneous Wire: Gold Wire with Gold Contacts}

The first system we consider is a homogeneous atomic wire of
gold. An atomistic diagram of one of the wires is shown in
Fig. \ref{fig1}. The atomic chain is made up of gold atoms with a
spacing of 2.4 \AA.  The chain is bonded over two gold tips,
each consisting of 13 gold atoms and both oriented in the (100)
direction.  The perpendicular bonding distance between the
chain and a tip is taken to be 2.0 \AA.  Each tip is then
bonded to a layer of 16 gold atoms which forms the unit cell
for the contact.  The gold atoms along the chain and the four
gold atoms in each tip to which the chain is attached are
modeled using the 6\,s6\,p5\,d orbitals of gold.  The remaining
gold atoms in the tips and those in the leads are modeled using
a 6\,s orbital. We consider chains ranging in length
from 3 to 7 gold
atoms.

\begin{figure}[ht]
\includegraphics[bb = 0 0 640 800, width = 0.85\textwidth,clip]{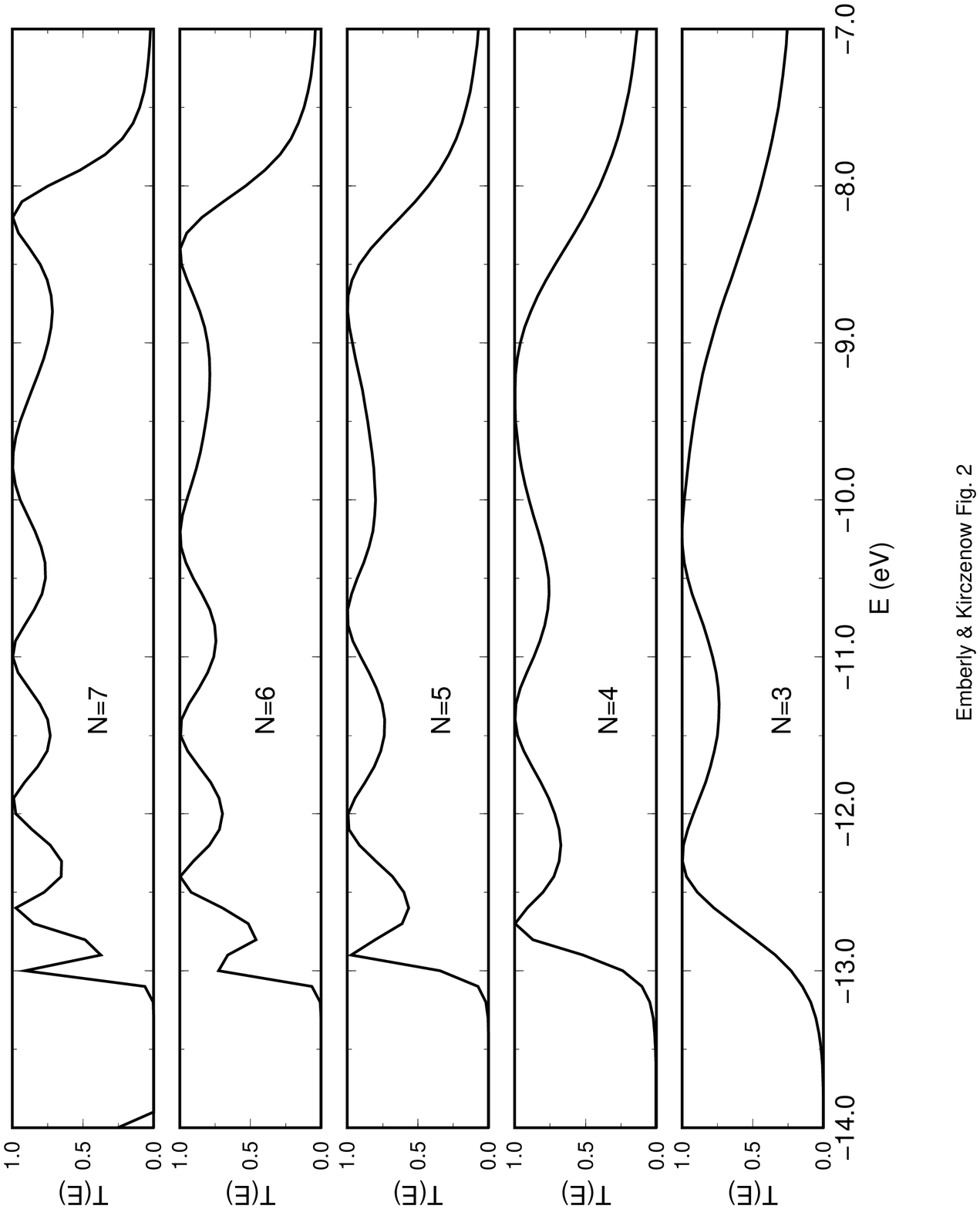}
\begin{center}
\caption{Transmission probability diagrams for $N$ gold atom
chains, where $N = 3\ldots 7$.}
\end{center}
\label{fig2}
\end{figure}

The transmission probabilities for the five different chain
lengths are shown in Fig. \ref{fig2}.  As expected there is a
different number of resonances for each chain length. For the
energies shown, only a 6\,s mode is supported by the chain and
hence the maximum value that the transmission probability can
take is 1 (ignoring spin). If there were more modes
which could propagate
along the chain the multi-channel Landauer
transmission probability could be greater
than 1. Since the gold atomic wires are a homogeneous system
the length corresponds approximately to the contact-to-contact
spacing which is roughly $(N+1)a$ where $N$ is the number of
atoms forming the chain and $a=2.4$\AA. So an $N$ atom chain
should display $N$ resonances and this is seen. (One might
argue that there should be $N+1$ resonances, but the
wavefunction that corresponds to this $n$ value has a node on
each atomic site and hence is unphysical). The $N^{th}$ or
last resonance of each chain (which is not shown) has an
energy that lies between -6.5 eV and -5.5 eV and this overlaps
some 6\,p modes which also contribute to the transmission and
obscure this resonance.

\begin{figure}[ht]
\includegraphics[bb = 0 0 640 800, width = 0.85\textwidth,clip]{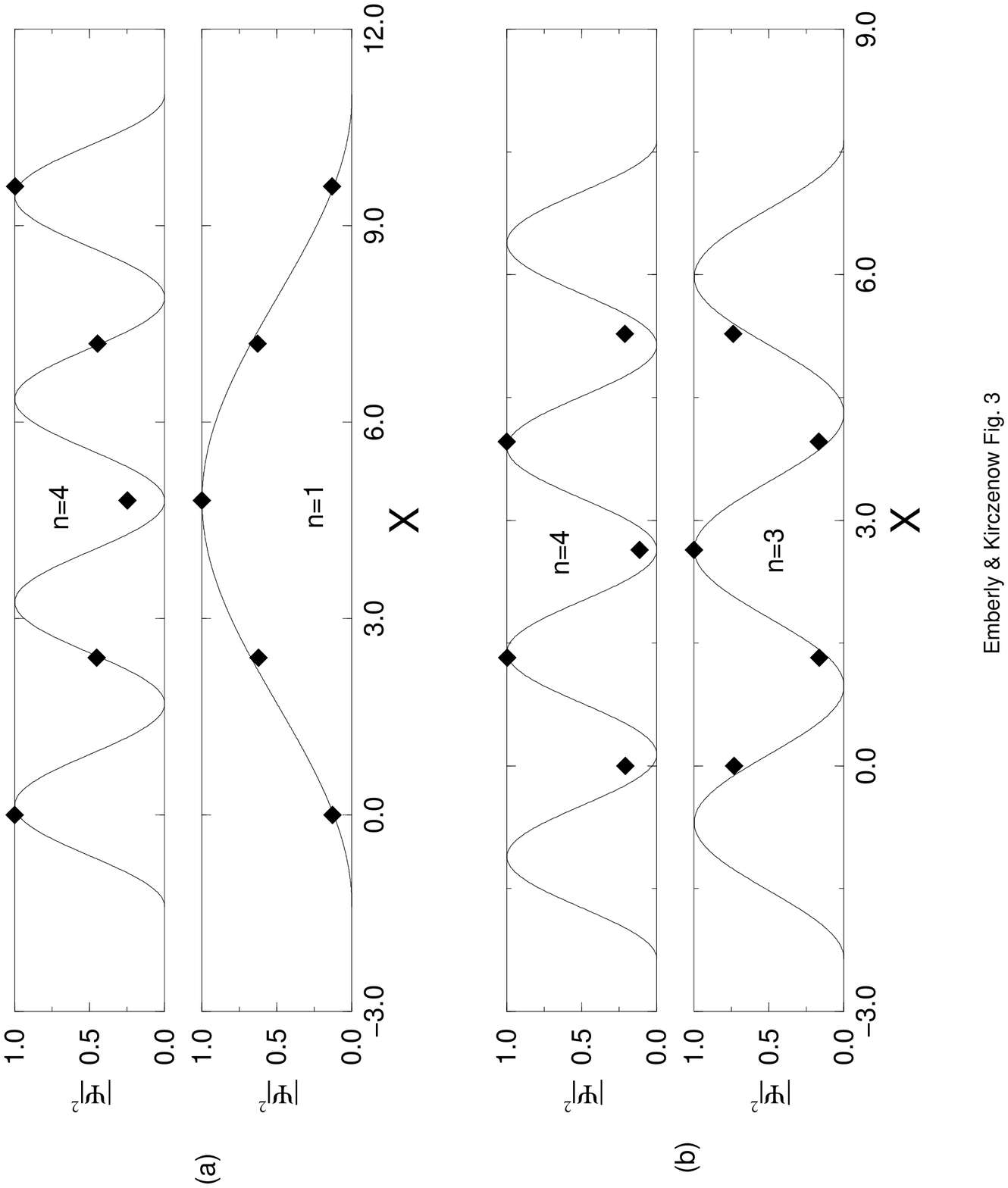}
\begin{center}
\caption{(a) Average squared wavefunctions (diamonds) for the $n=1,4$
transmission probability resonances of the 5 atom gold chain.
The corresponding normal modes, $\sin^2(n\pi (x-x_L)/L)$, with
$n=1,4$ are also plotted. The variable X measures the
length along the chain and is in \AA.
(b) Average squared wavefunctions for the $n=3,4$
transmission probability resonances of the 5 atom carbon chain.
The corresponding normal modes are shown for $n=3,4$.}
\end{center}
\label{fig3}
\end{figure}

\begin{figure}[ht]
\includegraphics[bb = 0 0 640 800, width = 0.85\textwidth,clip]{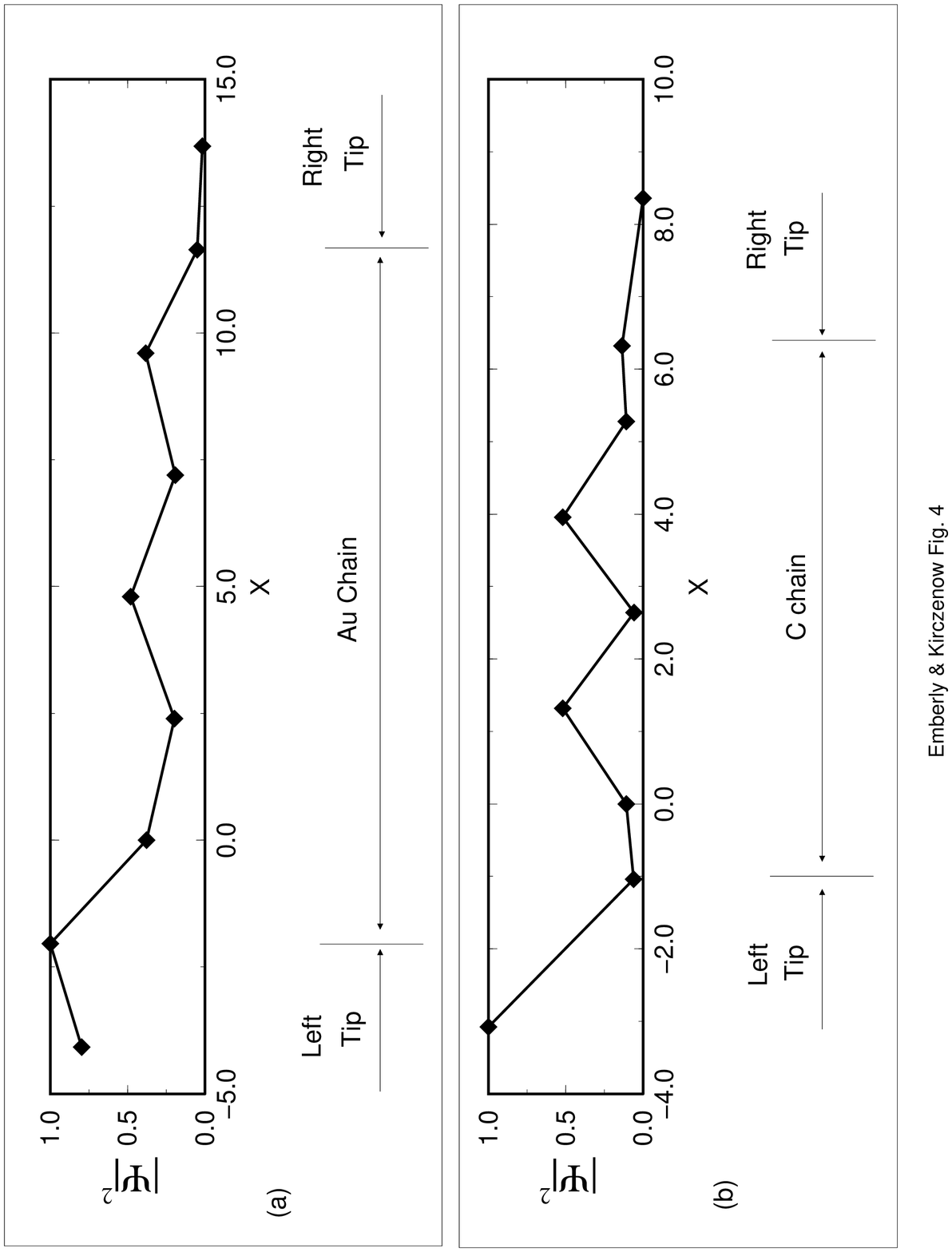}
\begin{center}
\caption{(a) The behaviour of the averaged squared wavefunction in the gold 
tips and the gold wire for the $n=3$ mode of the 5 gold atom
chain.
(b) The behaviour of the averaged squared wavefunction in the gold tips 
and the carbon wire for the $n=4$ mode of the 5 carbon atom
chain.}
\end{center}
\label{fig4}
\end{figure}

At each resonance the electron wavefunction is a standing wave
along the 6\,s orbital backbone of the chain.  The squared
modulus should behave
 as $\sin^2(n\Pi (x-x_L)/L)$ where $x$ is the
position coordinate along the wire ($x=0$ marks the first chain
atom), $x_L$ is the leftmost position where the wavefunction is
zero (this may extend beyond the chain into the tips) and $n$
labels the mode.  The wavefunction (calculated using our
tight-binding model) for the 5 atom chain is
plotted for the $n=1$ and $n=4$ resonance together with the corresponding 
ideal sinusoidal standing waveform in
Fig. \ref{fig3}a.  The graphs clearly indicate the formation of
standing waves along the chain.  The length $L$ which best fits
the data is around 13\AA which is just slightly shorter than
the 13.6\AA contact-to-contact spacing for this chain.  In
Fig. \ref{fig4}a the behaviour of the wavefunction in the gold
tips for the $n=3$ mode is depicted for the 5 atom chain.
The atoms in the tips fall outside the length of the "pipe" and
thus are not part of the standing wave formation within the
pipe. The wavefunction is consistently larger within the left
tip than on the right as is to be expected for a wave incident
on the wire from the left contact.  (Note that the squared
wavefunction that is depicted is a qualitative result and
should not be compared to the transmission results. What is
shown is the result of adding together the squared amplitude of
the wavefunction for each incident channel.  Thus it is the
``average'' wavefunction along the atomic chain.)

The conductance of these chains is given by the Landauer
formula, $G = G_0 T$.  The transmission probability that
enters the Landauer expression is evaluated at the source
Fermi energy.  For our gold leads the Fermi energy is
approximately -10 eV.
Fig. \ref{fig6}a shows the conductance as a
function of the number of atoms in the wire at this Fermi
energy. In qualitative agreement with
experiment\cite{Yan98} the conductance is nearly unity and
varies between 0.81 $G_o$ and 0.99 $G_o$. The fluctuations are
relatively small because the peak to valley ratio of the
transmission probability near the Fermi energy is near
unity for the different gold chains.

\section{Heterogeneous Wire: Carbon Wire Results}
The second structure that we consider is a heterogeneous system
consisting of a carbon atomic wire and gold contacts.  The
atomic system is similar to that depicted in Fig. \ref{fig1}.
The carbon chain spacing is taken to be 1.32\AA.  Again it is
bonded over 13-atom gold tips oriented in the (100) direction.
The perpendicular bond distance between the chain and a tip was
chosen to be 1\AA. The four gold atoms that bond to the carbon
wire on each tip were simulated using their 6\,s6\,p5\,d
orbitals while the other gold atoms were simulated using just
their 6\,s orbital. (We remark that the results obtained
modelling all of the gold atoms in this system with only
6s orbitals were qualitatively similar.)
The carbon atoms were simulated using
their 2\,s and 2\,p orbitals. Chains consisting of 4 to 10
carbon atoms were simulated.

\begin{figure}[ht]
\includegraphics[bb = 0 0 640 800, width = 0.85\textwidth,clip]{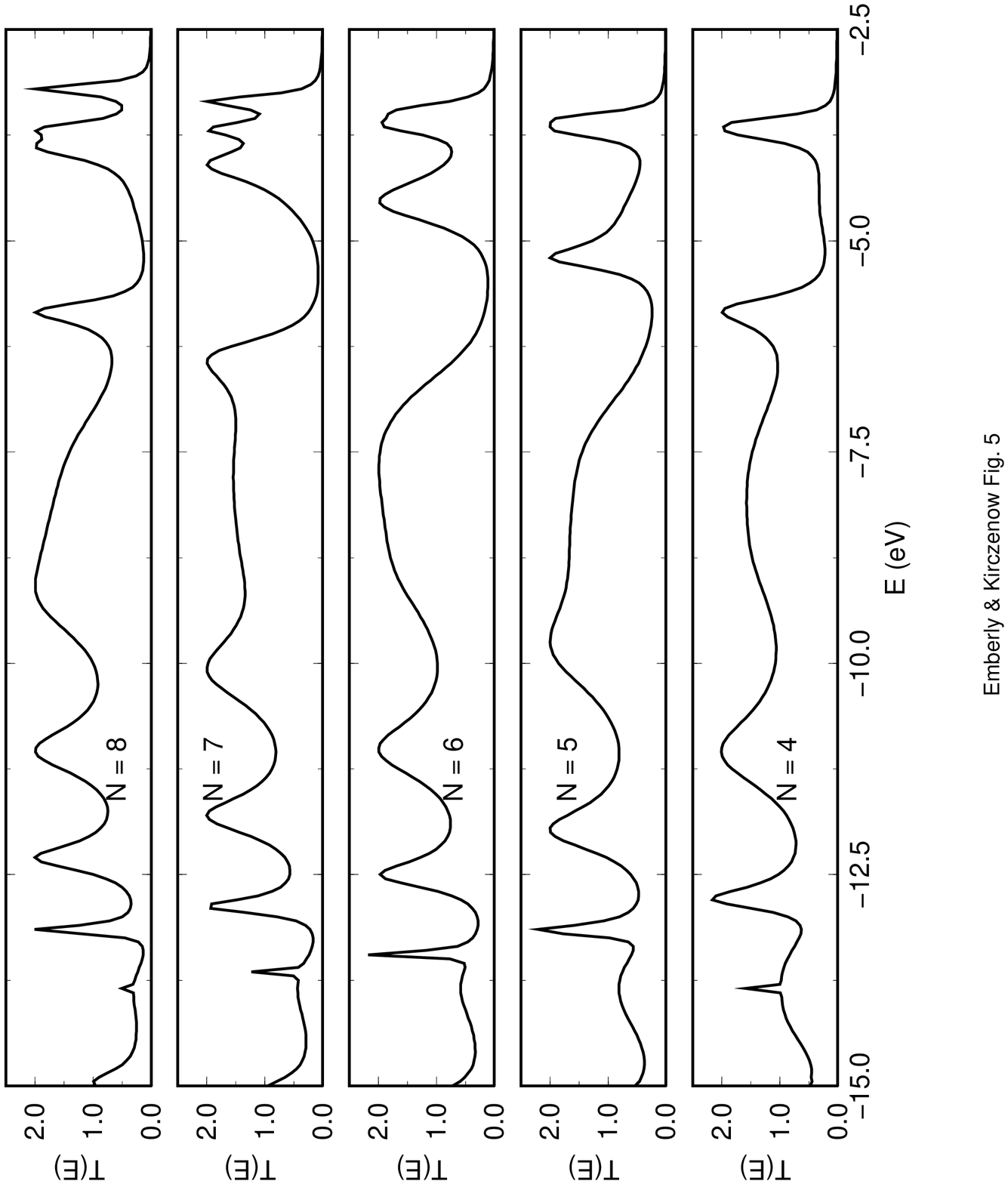}
\begin{center}
\caption{Transmission probability diagrams for $N$ carbon
atom chains bonded to gold contacts, where $N = 4,\ldots, 8$.}
\end{center}
\label{fig5}
\end{figure}

\begin{figure}[ht]
\includegraphics[bb = 0 0 640 800, width = 0.85\textwidth,clip]{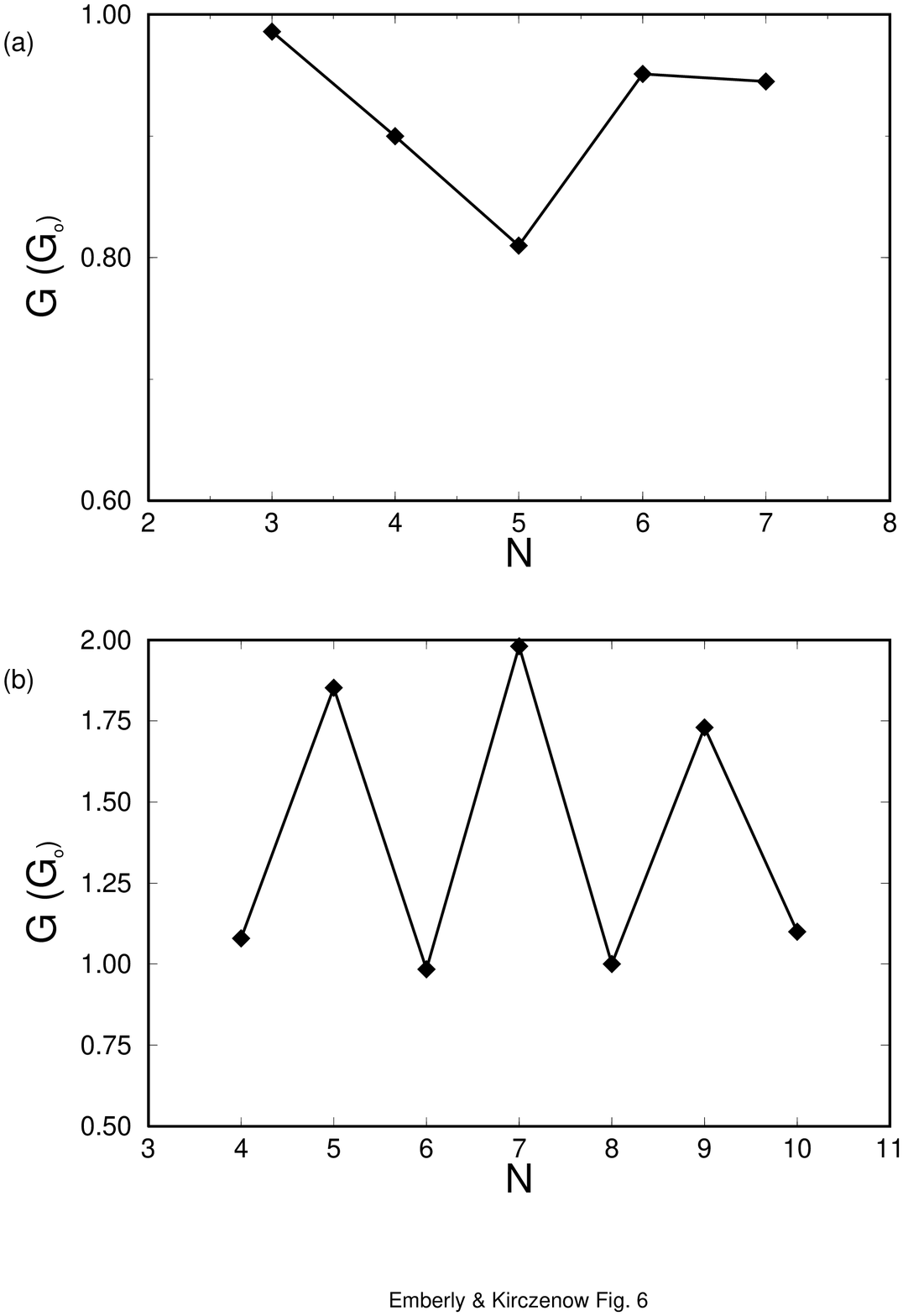}
\begin{center}
\caption{(a) Conductance of $N$ gold atom chains, where $N =
3,\ldots,7$. (1 $G_o = 2 e^2/h)$. The Fermi energy of the
contacts was chosen to be -10.0 eV.
(b) Conductance of $N$ carbon atom chains bonded to gold
leads, where $N = 4,\ldots, 10$. The Fermi energy of the contacts
was chosen to be -10.0 eV.}
\end{center}
\label{fig6}
\end{figure}

The transmission probabilities for chains numbering 4 through 8
are shown in Fig. \ref{fig5}.  
The magnitude is greater than that found in the gold chains with a maximum 
value of near 2.0.  This is because for these energies the electrons are 
propagating along the $\pi$ backbone of the carbon wire.  If the wire is 
oriented along the $z$ direction there are two independent modes ($p_x$ 
and $p_y$) along which the electrons can propagate.  Thus the 
transmission can be a maximum of 2.0 (ignoring spin).  In those regions 
where it exceeds 2.0, an extra $\sigma$ mode is present.  The number of 
resonances for a
given numbered carbon chain is also greater than that of the
corresponding gold chain.  
For heterogeneous 
systems the chemistry at the interface between the atomic wire and the
metallic contacts effectively alters the length of the 1D
conducting channel. For example the 4 atom carbon chain
displays 6 resonances which shows that the interaction is such
that it is effectively lengthened and behaves like a 6 atom
chain.

To determine what the effective length for one of these carbon
chains is we examine the electron's wavefunction.  The
wavefunctions for two different resonances of the 5 atom carbon
chain are plotted in Fig. \ref{fig3}b.  The corresponding normal
modes are also plotted where $L$ was chosen so as to give a
best fit to all the curves. The value of $L$ was determined to
be around 10.0\AA which is significantly longer than the
contact-to-contact distance of 7.28\AA.  From this we can
estimate that the interface region extends about 1.36\AA into
the gold tip.  So in this system the standing waves are forming
from within the gold tips rather than at their surfaces as was
the case for the gold atom wires. The behaviour of the
wavefunction in the tips is depicted in Fig. \ref{fig4}b.  It is
seen that the standing wave continues onto the first layer of
each tip.  However, the wavefunction on the second atomic layer
of each tip lies outside the length $L$ and resembles the
behavior of the wavefunction within the tips of the gold atomic
wire. It is seen that the wavefunction on the left tip is
greater than that on the right. Again, we emphasize
that we are using the ``average'' wavefunction to perform this
analysis.

The conductance of these carbon chains was found using the
same source Fermi energy of -10 eV.  Conductance oscillations
are found for this choice as the number of carbon atoms
in the chain is varied, with odd numbered chains possessing
a higher conductance than even ones (shown in
Fig. \ref{fig6}b).  This is consistent with other theoretical
work which has also found conductance oscillations as a
function of carbon wire length although the carbon wire and
contacts were modelled in a different way in that
work.\cite{Lang98} The oscillation has quite a strong signature
with the conductance varying between 0.9 $G_o$ and 1.8 $G_o$.
For the odd numbered chains the conductance is large because
the Fermi energy corresponds to a standing wave resonance,
whereas the even numbered chains are off resonance.  The
standing wave interpretation seems to be a reasonable physical
interpretation for the oscillating conductance found here and
in a previous report.\cite{Lang98}

\section{Conclusions}
We have shown that electron transport through an atomic wire is
consistent with the interpretation that electron standing waves
are formed in the atomic wire which bridges metallic
contacts. They arise because of the interference of reflected
electron waves at the interface between the atomic chain and
the metallic tips. For the gold tips the effective length of
the constriction was roughly the contact-to-contact distance.
The conductance of the gold wires was also found to be near
unity for atomic wires ranging from 3 to 7 gold atoms, consistent
with experiment. The
behavior in the carbon wires was different.  The effective
length of the constriction was significantly larger signalling
that the formation of standing waves is sensitive to the
chemistry of the atomic wire-contact interface.  The
conductance was also found to oscillate as a function of chain
length and this can be understood in terms of the different
standing wave modes that are created in chains of differing
length. The observation of electron standing waves may be
facilitated by using atomic wires (as opposed to
semiconductor systems) since the constriction can not be
smooth, which will enhance the reflections.

This work was supported by NSERC.


\end{document}